# Unusual Pressure-Induced Periodic Lattice Distortion in SnSe$_2$

Jianjun Ying,[1, 2] Hari Paudyal,[3] Christoph Heil,[4, 5] Xiao-Jia Chen,[6] Viktor V. Struzhkin,[1] and Elena R. Margine[3, *]

[1]*Geophysical Laboratory, Carnegie Institution of Washington, Washington, D.C. 20015, USA*
[2]*HPCAT, Geophysical Laboratory, Carnegie Institution of Washington, Argonne, Illinois 60439, USA*
[3]*Department of Physics, Applied Physics, and Astronomy,*
*Binghamton University-SUNY, Binghamton, New York 13902, USA*
[4]*Department of Materials, University of Oxford, Parks Road, Oxford OX1 3PH, United Kingdom*
[5]*Institute of Theoretical and Computational Physics,*
*Graz University of Technology, NAWI Graz, 8010 Graz, Austria*
[6]*Center for High Pressure Science and Technology Advanced Research, Shanghai 201203, China*
(Dated: July 6, 2018)

We performed high pressure x-ray diffraction (XRD), Raman, and transport measurements combined with first-principles calculations to investigate the behavior of tin diselenide (SnSe$_2$) under compression. The obtained single-crystal XRD data indicate the formation of a $(1/3, 1/3, 0)$-type superlattice above 17 GPa. According to our density functional theory results, the pressure-induced transition to the commensurate periodic lattice distortion (PLD) phase is due to the combined effect of strong Fermi surface nesting and electron-phonon coupling at a momentum wave vector $\mathbf{q} = (1/3, 1/3, 0)$. In contrast, similar PLD transitions associated with charge density wave (CDW) orderings in transition metal dichalcogenides (TMDs) do not involve significant Fermi surface nesting. The discovered pressure-induced PLD is quite remarkable, as pressure usually suppresses CDW phases in related materials. Our findings, therefore, provide new playgrounds to study the intricate mechanisms governing the emergence of PLD in TMD-related materials.

Transition metal dichalcogenides (TMDs) and similar layered materials have attracted significant attention in recent years due to their novel electronic and optical properties [1, 2]. Bulk TMDs exhibit rich physics and have a promising potential for technological applications [3, 4]. Coexisting charge density wave (CDW) order and superconductivity were often found at low temperatures in metallic TMDs, providing an ideal platform to investigate the interplay of these quantum phases [3, 5]. Various commensurate modulated superstructures related to the CDW have been observed in TMDs, such as a $\sqrt{13} \times \sqrt{13} \times 1$ superlattice in $1T$-Ta$X_2$($X$=S, Se), a $3 \times 3 \times 1$ in $2H$-$M$Se$_2$ ($M$=Nb, Ta) and a $2 \times 2 \times 2$ superlattice in $1T$-TiSe$_2$ [6–8], and their origins have been to a large part attributed to strong electron-phonon coupling (EPC) for particular phonon modes and wave vectors [9–11]. Many novel physical properties were recently discovered in TMDs: large positive magnetoresistance was observed at low temperatures in WTe$_2$ [12], and both WTe$_2$ and MoTe$_2$ were suggested as type II Weyl semimetal candidate materials [13, 14]. Insulator-metal transitions and superconductivity were also observed in compressed TMDs [4, 15–18], raising expectations of novel phenomena in related new materials under high-pressure conditions.

Similar to related TMDs, SnSe$_2$ is a semiconductor with a bulk band gap of about 1.0 eV [19], and few-layer sheets of SnSe$_2$ were suggested to have potential applications in electronic and optoelectronic devices [20]. It has been proposed that SnSe$_2$ can rapidly and reversibly switch from amorphous to crystalline under laser heating, which leads to significant changes in the optical reflectivity; and thus, provides excellent perspective for data storage applications [21]. Recent theoretical calculations suggested that SnSe$_2$ is unstable above 20 GPa, and that it would probably decompose into Sn$_3$Se$_4$ and Se at high pressures [22, 23]. However, there is no experimental confirmation of the proposed decomposition scenario in compressed SnSe$_2$. In this study, we find that SnSe$_2$ can, in fact, be stabilized above ∼17 GPa by a periodic lattice distortion (PLD) rather than undergoing a decomposition. This PLD is similar to the CDW order observed in TMDs; however, their origins are quite different, as will be discussed in this Letter.

We performed high-pressure single crystal diffraction measurements on the hexagonal SnSe$_2$ structure ($H1$ phase) (see Supplemental Material [24]). The low-pressure x-ray diffraction (XRD) patterns are consistent with the previously reported data at ambient pressure [39]. When the pressure is increased above 17 GPa, diffraction patterns reveal superlattice reflections at $(1/3, 1/3, 0)$, as shown in Fig. 1. This clearly indicates a tripled unit cell above 17 GPa ($H2$ phase), and the XRD patterns at high pressure can be indexed very well using the $\sqrt{3} \times \sqrt{3} \times 1$ superlattice (Supplemental Fig. S1 [24]). The pressure dependence of the lattice parameters $a$ and $c$ is shown in Fig. 1(c). As is evident from the figure, the normalized in-plane lattice parameter does not show any anomaly across the phase transition. However, the lattice parameter $c$ and, consequently, the volume per formula unit both have a kink around the phase transition. Taken together, these experimental findings show that the phase transition does not involve a change in the hexagonal lattice structure, but a modulation of the atomic positions as observed in many CDW phases.

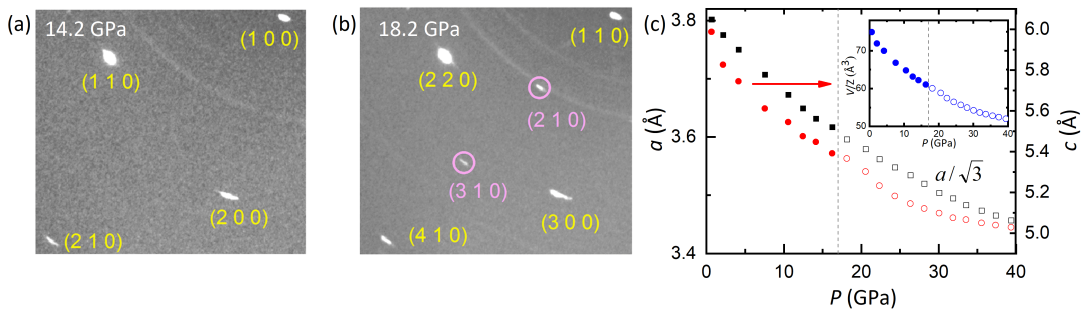

FIG. 1. XRD patterns below (a) and above (b) the phase transition. The sudden appearance of superlattice peaks [magenta circles in (b)] is indicative for a $\sqrt{3} \times \sqrt{3} \times 1$ superlattice formation. (c) The pressure dependence of the lattice parameters $a$ (black squares, left axis) and $c$ (red circles, right axis). The inset shows the pressure dependence of the volume per one formula unit ($V/Z$). For comparison with the $H1$ phase, $a$ in the $H2$ phase was rescaled by $1/\sqrt{3}$. Small kinks are observed in $c$ and $V/Z$ at 17 GPa.

In order to elucidate the observed phase transition, we performed first-principles dynamical stability calculations [24]. Based on the phase transformation revealed in the XRD pattern, we began by examining the phonon modes at the $\Gamma$ point in the $\sqrt{3} \times \sqrt{3} \times 1$ $H1$ superlattice. As the transition pressure is approached, we observe that the two lowest-energy phonon modes soften and exhibit imaginary frequencies above 18 GPa [40]. This is indicative of a lattice instability and in line with findings in metallic TMDs exhibiting a CDW phase [10, 11, 44, 45]. Furthermore, as shown in Fig. 2(a), the decrease in frequency is nonlinear with pressure, reflecting the strong changes in the electronic band structure that take place at the Fermi level (more details later in the manuscript). The squared frequency, on the other hand, exhibits a linear dependence with pressure in the transition region, a characteristic of a soft-mode phase transition described by Landau theory [46]. Incidentally, calculations of the phonon dispersion at 30 GPa in the three-atom $H1$ unit cell show that the lowest-energy vibrational mode has an imaginary frequency at the $K$ point in the Brillouin zone (BZ), consistent with the BZ folding of the $K$ point of the $1 \times 1 \times 1$ unit cell onto the $\Gamma$ point of the $\sqrt{3} \times \sqrt{3} \times 1$ supercell (Supplemental Fig. S2 [24]).

We further explored the full potential energy surface of the $H1$ $\sqrt{3} \times \sqrt{3} \times 1$ supercell for the soft phonon modes at 30 GPa. The atoms were displaced according to (i) the eigenvector of the $A_{2g}$ phonon ($H2-1$ phase), (ii) the eigenvector of the $A_{2u}$ phonon ($H2-2$ phase), and (iii) the combined displacements of the two phonon eigenvectors ($H2-3$ phases) [47]. The energy surface has a Mexican hat shape with many local minima within 1 meV/atom from each other (Supplemental Fig. S4(a) [24]). To find the equilibrium configuration for the distorted system, we picked six low-energy points on this surface for which we performed full geometrical optimization. All selected configurations gain in enthalpy with respect to the $H1$ phase and become virtually degenerate in enthalpy after relaxation (Supplemental Fig. S4(b) [24, 48]), yet belong to three different space groups [49] (Supplemental Table S1 and Fig. S5 [24]). Noticeably, the out-of-plane modulation in $H2-3$ is an intermediate state along the pathway that transforms $H2-1$ into $H2-2$. We will therefore, without loss of generality, concentrate further only on the $H2-1$ and $H2-2$ phases.

Comparing the enthalpies for the $H1$ and $H2$ structures as function of pressure reveals that the latter are energetically more favorable above 18 GPa, as shown in Fig. 2(b), in very good agreement with our experiments and phonon calculations. Further in depth analysis of the $H2$ derivatives confirmed that they remain degenerate in enthalpy within the $0-40$ GPa range considered, and the differences in their vibrational and electronic properties are negligible (Supplemental Figs. S2(c), S2(d), S6, and S7 [24]).

While our work demonstrates the stabilization of $H1$ derivatives obtained through atomic distortion, previous studies have pointed out the possibility of stable phases obtained through interlayer shifts. Namely, an alternative long-period stacking sequence with space group $R\bar{3}m$ [52] has been recently predicted to be 3.5 meV/atom higher in enthalpy at 0 GPa, but to stabilize with respect to $H1$ at 5 GPa [22]. Observation of such polymorphs in high-pressure experiments is not straightforward, as the kinetics of the transformation is likely defined by high-energy barriers of interlayer shifts. Our calculations with finer convergence settings, however, indicate that compared to $H1$, the $R\bar{3}m$ polymorph is virtually degenerate in enthalpy at 0 GPa, reaches a minimum relative enthalpy of $\sim$2.3 meV/atom at 6 GPa, and eventually destabilizes by $\sim$5.7 meV/atom at 25 GPa. We attribute this behavior to the fact that the application of pressure drastically changes the inter-layer distance, promoting the formation of covalent bonds between Se atoms in adjacent layers (Supplemental Fig. S5 [24]). Such changes in bonding promote a change in compressibility along the $c$ axis, as observed in both experimental

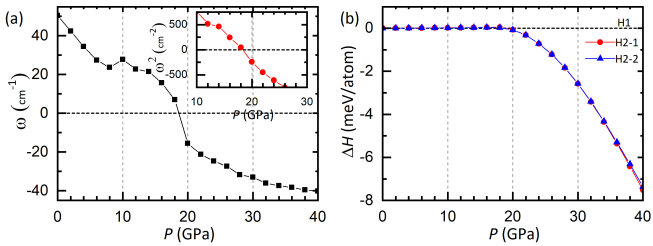

FIG. 2. (a) Calculated softening of the lowest-energy (degenerate) phonon modes, where imaginary phonon frequencies are shown as negative, illustrating the pressure-induced destabilization of the $H1$ phase. The inset shows the squared frequency of the modes. (b) Differences in enthalpies for the considered structures as function of pressure, where the $H1$ phase was chosen as reference.

and theoretical data (Supplemental Figs. S8 and S9 [24]).

High-pressure transport measurements were performed on SnSe$_2$ single crystals. At low pressures, the resistivity shows a semiconducting behavior, consistent with SnSe$_2$ being a narrow band semiconductor at 0 GPa (Supplemental Fig. S10 [24]). The resistivity decreases under pressure and the temperature dependence becomes flat around 8 GPa, signaling the closure of the band gap. The semiconductor-to-metal transition occurs in the $8 - 13$ GPa range, and a typical metallic behavior with much larger residual-resistance ratio is observed above 17 GPa. Similarly to the room temperature resistivity, the Hall coefficient decreases gradually with increasing pressure and displays a kink at ~15 GPa, around the $H1$-to-$H2$ phase transition [53]. These complementary results provide evidence for the Fermi surface (FS) reconstruction across the phase transition.

To investigate the metallization from the theoretical side, we carried out band structure calculations at various pressures. As illustrated in Supplemental Fig. S11 [24], the top of the valence band of $H1$ structure at 0 GPa consists of states with Se $p_z$ and $p_{xy}$ character. The conduction band minimum is formed by Se $p_z$ and $p_{xy}$ orbitals, which hybridize strongly with Sn $s$ orbitals [54]. With increasing pressure, the conduction Se $p_{xy}$ band shifts down and touches the valence Se $p_z$ band at $5 - 6$ GPa, closing the band gap. This pressure-induced metallization is consistent with our resistivity measurements, although the predicted semiconductor-to-metal transition pressure is smaller than the experimental value. This difference is partially due to the underestimation of the band gap in the semilocal density functional theory approximation used in this study [19, 55, 56]. Band structure calculations employing screened hybrid functional indeed predict a semiconductor-to-metal transition at about 8 GPa [23], in better agreement with experiments. An additional source for the offset could arise from the discrepancy between the calculated and experimental lattice constants in the $5 - 20$ GPa range. Since NaCl provides a slightly non hydrostatic pressure environment [24], we computed the band structure of $H1$ phase by compressing the $c$ axis, as done experimentally. As in $2H$-MoTe$_2$ [57], the out-of-plane compression favors the semiconductor-to-metal transition at a lower pressure; nevertheless, the effect is small (Supplemental Fig. S12 [24]), and the deviation from hydrostatic pressure should have a minor effect on the semiconductor-to-metal transition pressure.

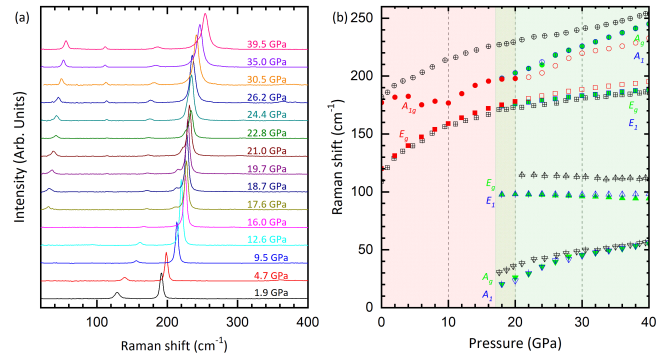

FIG. 3. (a) Selected Raman spectra for various pressures. New modes appear above 17 GPa, indicating the $H1$-to-$H2$ phase transition. (b) Calculated and measured frequency dependence of the Raman-active modes as a function of pressure. The theoretical results are shown as red symbols ($H1$), green symbols ($H2-1$), and blue symbols ($H2-2$). For $H1$, the data before and after the phase transition are shown as filled and open symbols, respectively. The experimental data are shown as black symbols.

To shed more light on the pressure-induced structural modulations, we performed high-pressure Raman measurements, as shown in Fig. 3(a). The low-pressure spectrum has two peaks around 120 cm$^{-1}$ ($E_g$) and 190 cm$^{-1}$ ($A_{1g}$), respectively [58], in agreement to previous results [59]. The two peaks shift to higher frequency as the pressure increases. Above 17 GPa, two new peaks appear, indicating the lowering of crystalline symmetry, consistent with our XRD results.

In Figure 3(b), we compare the frequencies of the calculated and observed Raman-active modes [60]. The pressure dependence of the high-energy $E$-type mode is in excellent agreement with the experimental results, clearly showing the structural phase transition in the $17-20$ GPa range. The high-energy $A$-type mode, while it is appreciably underestimated compared to experiment, does exhibit a similar trend, as discussed in the Supplemental Material [24]. After the phase transition, the in-plane unit cell triples in size and two additional Raman-active modes appear, consistent with the Raman measurements. The low-energy $A$-type mode in the $H2$ phases hardens with increasing pressure, and follows closely the experimental data [61]. The medium-energy $E$-type mode (~100 cm$^{-1}$) is slightly downshifted compared to experiment, but displays a similar pressure dependence [62].

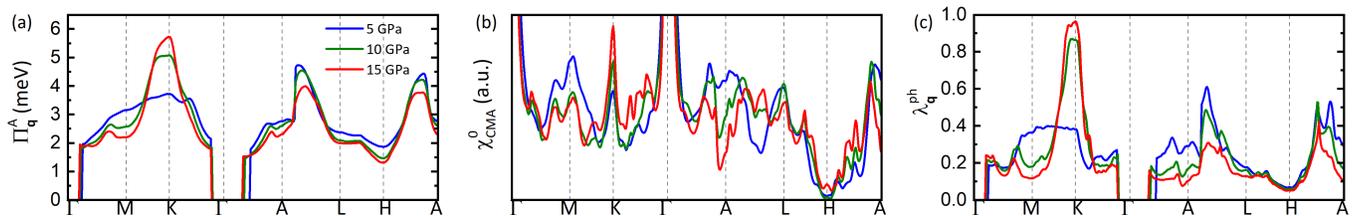

FIG. 4. (a) The FS contribution to the adiabatic phonon self-energy, (b) the static bare susceptibility in the constant matrix approximation, and (c) the EPC strength for the soft phonon mode in the three-atom $H1$ structure at various pressures [67]. These results provide evidence that both FS nesting and electron-phonon interactions play crucial parts in the observed phase transition of SnSe$_2$.

A comparison between the electronic structures of the $H1$ and $H2$ compounds reveals that the phase transition leads to significant changes. In particular, we find avoided crossings near the Fermi level along $\Gamma$-$M'$, $\Gamma$-$K'$, and $\Gamma$-$A$ directions in the $H2$ phases (Supplemental Figs. S6 and S7 [24]). These changes are due to structural displacements originating from a phonon mode (Supplemental Fig. S13 [24]), pointing towards a sizable EPC of that phonon with electronic states near the Fermi level [63, 64]. The deformation of the band structure is accompanied by a suppression of large parts of the FS (Supplemental Fig. S14 [24]), and an increasing removal of electronic weight at the Fermi level with increasing pressure (Supplemental Fig. S10(d) [24]). Similar characteristics have been observed in previous theoretical studies of CDW order in metallic TMDs [10, 11, 44, 45].

Usually, the appearance of a superlattice is the signature of a CDW transition in low-dimensional systems, giving rise to the question, of whether the transitions are driven primarily by an instability of the electronic subsystem, the vibrational subsystem, or both. In metallic TMDs it has been shown that the wave vector dependence of the EPC drives the CDW formation, while the FS nesting has been found to play a minor role [9–11, 44, 45]. In particular, the breaking of electronic degeneracies by a phonon-modulated lattice distortion, the reduction of the density of states at the Fermi level, and the softening of a low-energy phonon are key signatures of a momentum-dependent electron-phonon coupling CDW instability.

To understand the phonon softening with pressure and to investigate the mechanisms behind the phase transition, we calculated the FS contribution to the adiabatic phonon self-energy $\Pi^A_{\mathbf{q},\nu}$ [65] for the soft phonon mode of the three-atom $H1$ phase [66]. As shown in Figure 4(a), $\Pi^A_{\mathbf{q},\nu}$ increases dramatically with pressure at the $K$ point, explaining the observed phonon softening and instability towards a transition to the $H2$ phase. In order to separate the purely electronic effects from that of the electron-phonon interaction, we calculated the bare susceptibility $\chi^0_{CMA}$ in the constant matrix approximation [71] and the EPC $\lambda^{ph}_{\mathbf{q},\nu}$ [see Figs. 4(b) and 4(c)]. Both $\chi^0_{CMA}$ and $\lambda^{ph}_{\mathbf{q},\nu}$ show a strong increase at the $K$ point with increasing pressure, indicating that both FS nesting and electron-phonon interactions play crucial parts in the observed phase transition of SnSe$_2$. We find that the largest contributions to the FS nesting come from aligning the $\Gamma$-centered surface with the $K$-centered one (Supplemental Fig. S14(a) [24]). This observation is in contrast to findings in metallic TMDs, where FS nesting was found to play a less significant role in the transition, as mentioned above [10, 11, 45, 72].

The application of pressure would usually suppress the CDW in metallic TMDs [3, 5]. Our result shows an example of the opposite situation: the pressure can actually induce a PLD (and possibly CDW) in a semiconducting TMD-related material. Although pressure-induced metallization was also observed in many semiconducting TMDs, no evidence of pressure-induced PLD (or CDW) has been observed so far, according to our knowledge in TMD-related materials. We expect that future research will show if the PLD observed here is related to a true CDW quantum state [73].

In conclusion, we find a pressure-induced phase transition in SnSe$_2$, which we investigated in detail by means of XRD, Raman and transport measurements, as well as first-principles calculations. Our experiments and calculations are overall in very good agreement with each other. We have shown that following the phonon instability naturally leads to the transition to a supercell structure, whose electronic and vibrational properties match our experimental findings. We further provide evidence that the observed phase transition is due to the combined effect of strong FS nesting and EPC at a momentum wave vector $\mathbf{q} = (1/3, 1/3, 0)$. To our knowledge, the formation of a $(1/3, 1/3, 0)$-type superlattice with application of pressure was never observed in TMD-related materials. Our discovery of pressure-induced structural modulation, therefore, offers exciting new possibilities and provides a new playground to study PLD and possible CDW phases in TMD-related materials.

We thank Changyong Park for help with the XRD experiment. High-pressure experiments were supported by DOE/BES under Contract No. DE-FG02-99ER45775.

We acknowledge the usage of PPMS being supported by Energy Frontier Research in Extreme Environments Center (EFree), an Energy Frontier Research Center funded by the U.S. Department of Energy, Office of Science under Award Number DE-SC0001057. Portions of this work were performed at HPCAT (Sector 16), Advanced Photon Source (APS), Argonne National Laboratory. HPCAT operation is supported by DOE-NNSA under Award No. DE-NA0001974, with partial instrumentation funding by NSF. The Advanced Photon Source is a U.S. Department of Energy (DOE) Office of Science User Facility operated for the DOE Office of Science by Argonne National Laboratory under Contract No. DE-AC02-06CH11357. H.P. and E.R.M. acknowledge the NSF support (Award No. OAC-1740263). C.H. acknowledges support by the Austrian Science Fund (FWF) Project No. J3806-N36 and the Vienna Science Cluster.

---

# Unusual pressure-induced periodic lattice distortion in SnSe₂

Jianjun Ying,[1, 2] Hari Paudyal,[3] Christoph Heil,[4, 5] Xiao-Jia Chen,[6] Viktor V. Struzhkin,[1] and Elena R. Margine[3, *]

[1]*Geophysical Laboratory, Carnegie Institution of Washington, Washington, DC 20015, USA*
[2]*HPCAT, Geophysical Laboratory, Carnegie Institution of Washington, Argonne, Illinois 60439, USA*
[3]*Department of Physics, Applied Physics, and Astronomy,
Binghamton University-SUNY, Binghamton, New York 13902, USA*
[4]*Department of Materials, University of Oxford, Parks Road, Oxford OX1 3PH, United Kingdom*
[5]*Institute of Theoretical and Computational Physics,
Graz University of Technology, NAWI Graz, 8010 Graz, Austria*
[6]*Center for High Pressure Science and Technology Advanced Research, Shanghai 201203, China*
(Dated: July 13, 2018)

### High-pressure experiments

SnSe₂ single crystals were obtained from the commercial company HQ Graphene [1]. Diamond anvils with 300 $\mu m$ culet and c-BN+epoxy gasket with sample chambers having diameter of 100 $\mu m$ were used for transport measurements. The single crystal of SnSe₂ was cut to the dimensions of 60 $\mu m \times$ 60 $\mu m \times$ 10 $\mu m$. NaCl was used as a pressure transmitting medium and the pressure was calibrated by using the shift of ruby fluorescence at room temperature. During transport measurements, the pressure was applied at room temperature using the miniature diamond anvil cell [2]. The resistivity and Hall coefficient were measured using the Quantum Design PPMS-9. The high-pressure single crystal diffraction was carried out using the angle-dispersive X-ray diffraction facility at 16-BMD beamline of the High-Pressure Collaborative Access Team (HPCAT) at the Advanced Photon Source at the Argonne National Laboratory. A symmetric diamond anvil cell (DAC) with a pair of 300 $\mu m$ culet size anvils was used to generate pressure and Neon was loaded as a pressure transmitting medium. The diamond anvils with 300 $\mu m$ culets were used for the high pressure Raman measurements with incident laser wavelength of 488 nm at the Geophysical Laboratory. Neon was also used as a pressure transmitting medium in the Raman experiments. All samples were loaded in DACs with the *c*-axis along the applied pressure direction. Both Neon and NaCl provide a quasi-hydrostatic pressure environment below 40 GPa. We estimated the pressure variation to be within 2 GPa in our low-temperature transport experiments [2]. All the XRD and Raman measurements were performed at room temperature. The experimental error of the pressure measured at room temperature is within 0.3 GPa.

### First-principles calculations

The stability of the SnSe₂ under pressure was examined at the density functional theory (DFT) level with VASP [3, 4]. We employed projector augmented wave (PAW) potentials [5] with Perdew-Burke-Ernzerhof (PBE) exchange-correlation functional [6] in the generalized gradient approximation [7], where the Se $4s^24p^4$ and Sn $4d^{10}5s^25p^2$ orbitals were included as valence electrons. At ambient pressure, SnSe₂ crystallizes in the hexagonal, close-packed CdI₂-type structure with space group P$\bar{3}$m1 (no. 164). The unit cell contains three atoms, where every Sn atom occupies the center of an octahedron formed by six Se atoms. We refer to this structure as H1 phase, and all calculations, unless otherwise specified, were performed for the nine-atom $\sqrt{3} \times \sqrt{3} \times 1$ superlattice. To properly treat the long-range dispersive interactions, we used the non-local van der Waals (vdW) density functional optB86b-vdW [8, 9]. A Γ-centered 12 × 12 × 12 Monkhorst-Pack **k**-mesh [10], a plane-wave energy cutoff of 500 eV, a smearing value of 0.1 eV, and a convergence criterion of $10^{-6}$ eV for the self-consistent energy were used in the atomic and electronic structure calculations. The atomic positions and lattice constants were relaxed until the maximum force on each atom was less than 0.003 eV/Å and the stresses were converged within 0.1 kB. We obtained phonon dispersions with PHONOPY [11] using the force constants calculated in VASP within density functional perturbation theory. For this, we used a Γ-centered 6 × 6 × 6 **k**-mesh and a 2 × 2 × 2 expansion of the $\sqrt{3} \times \sqrt{3} \times 1$ superlattice. The structural parameters of all phases considered are provided in Table S1.

The calculation of the electron-phonon interactions, bare susceptibilities and phonon self-energies were obtained with the EPW code [12, 13] of the QUANTUM ESPRESSO distribution [14], in conjunction with the Wannier90 library [15]. We used PBE norm-conserving pseudopotentials [16] with a plane-wave kinetic energy cutoff of 800 eV. The electronic eigenvalues and phonon dynamical matrices of the three-atom H1 structure were calculated with QUANTUM ESPRESSO on coarse grids of $24 \times 24 \times 16$ **k**-points and $6 \times 6 \times 4$ **q**-points, respectively. The electron-phonon matrix elements were obtained using the Wannier-Fourier interpolation technique implemented in EPW. The interpolation was performed on a fine $300 \times 300 \times 200$ **k**-mesh along a high-symmetry path in **q**-space. Gaussian smearings of 25 meV and 0.05 meV were used for the electronic and vibrational properties evaluated with EPW.

### Irreducible representations of the optical zone-center phonons

The presence of inversion symmetry in H1 and H2-1 dictates that the zone-center lattice vibrations are either Raman or infrared active. On the other hand, some of the infrared active modes in H1 and H2-1 become both Raman and infrared active in H2-2 and H2-3 due to the lack of inversion symmetry in the latter cases. The irreducible representations [17, 18] of the optical zone-center phonons in the four SnSe$_2$ configurations are:

$$\Gamma_{\text{H1}} = A_{1g} + A_{2u} + E_u + E_u, \tag{1}$$

where $A_{1g}$ and $E_g$ are Raman active, and $A_{2u}$ and $E_u$ are infrared active

$$\Gamma_{\text{H2-1}} = 4A_g + 4A_u + 4{}^1E_g + 4{}^1E_u + 4{}^2E_g + 4{}^2E_u, \tag{2}$$

where $A_g$ and $E_g$ are Raman active, $A_u$ and $E_u$ are infrared active,

$$\Gamma_{\text{H2-2}} = 5A_1 + 3A_2 + 8E, \tag{3}$$

where $A_1$ and $E$ modes are both Raman and infrared active, and

$$\Gamma_{\text{H2-3}} = 8A + 8{}^1E + 8{}^2E, \tag{4}$$

where $A$, ${}^1E$, and ${}^2E$ modes are both Raman and infrared active. Note that there are six optical modes in the three-atom H1 phase, while 24 optical phonon modes in the nine-atom H2 phases.

### Evolution of the $A_{1g}$ mode under pressure

The discrepancies in the measured and calculated behaviors of the $A_{1g}$ mode frequencies as a function of pressure can be elucidated by the following analysis of the relevant electronic and vibrational features. The non-monotonic response of the $A_{1g}$ mode to pressure below 10 GPa could be related to the pressure-induced metallization of the initial H1 phase. Indeed, our frozen-phonon examination of the two Raman active modes in this pressure range indicates that $A_{1g}$ has the strongest coupling with the electronic states near the Fermi level, since the $A_{1g}$ modulation of atomic positions causes the largest shifts in the corresponding electronic states as shown in Fig. S13.

| SnSe₂ phase | Space group | a (Å) | c (Å) | Wyckoff positions | |
|---|---|---|---|---|---|
| | | | | Sn | Se |
| H1 (0 GPa) | 164 P$\bar{3}$m1 | 3.85917 | 6.11503 | 1a (0.00000, 0.00000, 0.00000) | 2d (0.33333, 0.66667, 0.73835) |
| H1 (30 GPa) | 164 P$\bar{3}$m1 | 3.54866 | 5.18142 | 1a (0.00000, 0.00000, 0.00000) | 2d (0.33333, 0.66667, 0.69058) |
| H2-1 | 147 P$\bar{3}$ | 6.14822 | 5.16299 | 2d (0.33333, 0.66667, 0.95662)  1a (0.00000, 0.00000, 0.00000) | 6g (0.31855, 0.96922, 0.30678) |
| H2-2 | 157 P31m | 6.14874 | 5.16215 | 2b (0.33333, 0.66667, 0.97529)  1a (0.00000, 0.00000, 0.04940) | 3c (0.35909, 0.00000, 0.30686)  3c (0.69353, 0.00000, 0.69314) |
| H2-3 | 143 P3 | 6.14962 | 5.15905 | 1a (0.00000, 0.00000, 0.95331)  1b (0.33333, 0.66667, 0.00243)  1c (0.66667, 0.33333, 0.04425) | 3d (0.01416, 0.31529, 0.30663)  3d (0.98460, 0.63458, 0.69337) |

TABLE S1. Space groups, lattice parameters, and Wyckoff positions of SnSe₂ structures fully relaxed at the DFT level. The space groups were found with the ISOTROPY software [19] interfaced with the MAISE package [20] using a tolerance of 0.01. The lattice parameters are given at 0 GPa and 30 GPa for H1 and at 30 GPa for the H2-1, H2-2, and H2-3 derivatives. The lattice parameters for H2-3 correspond to configuration number four in Fig. S4.

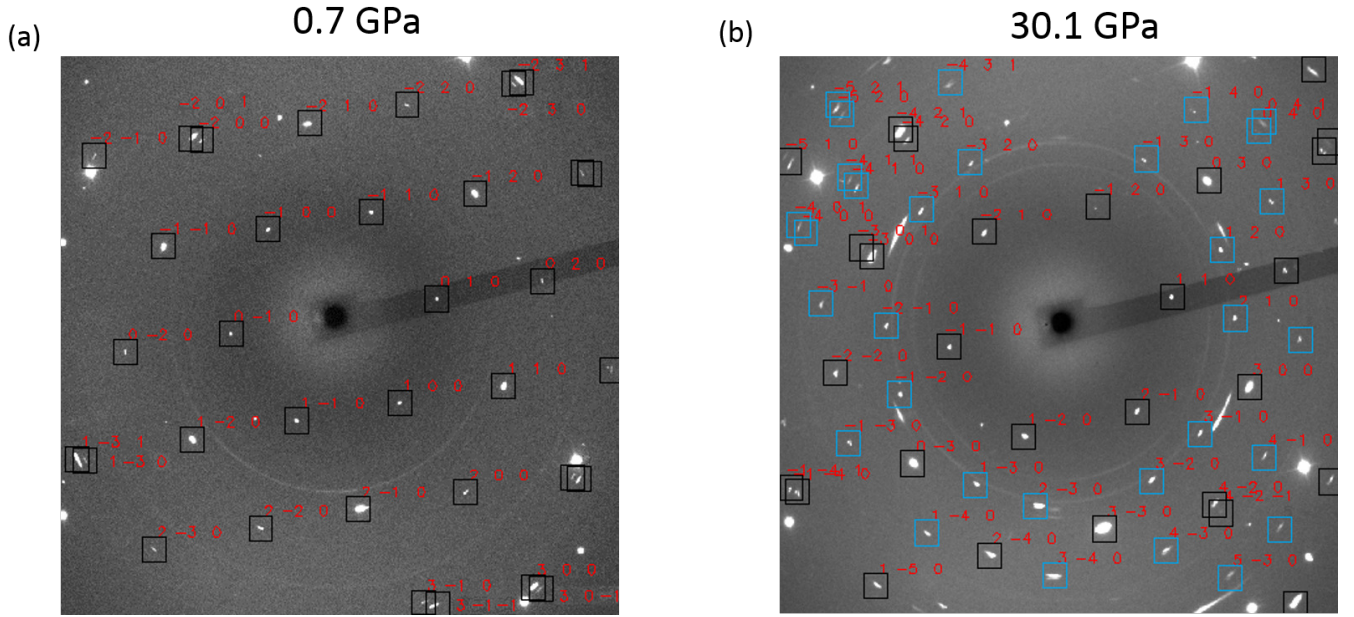

(a) 0.7 GPa  (b) 30.1 GPa

FIG. S1. Typical indexed XRD pattern before (a) and after (b) the phase transition. The blue squares correspond to the superlattice peaks in the H2 phase.

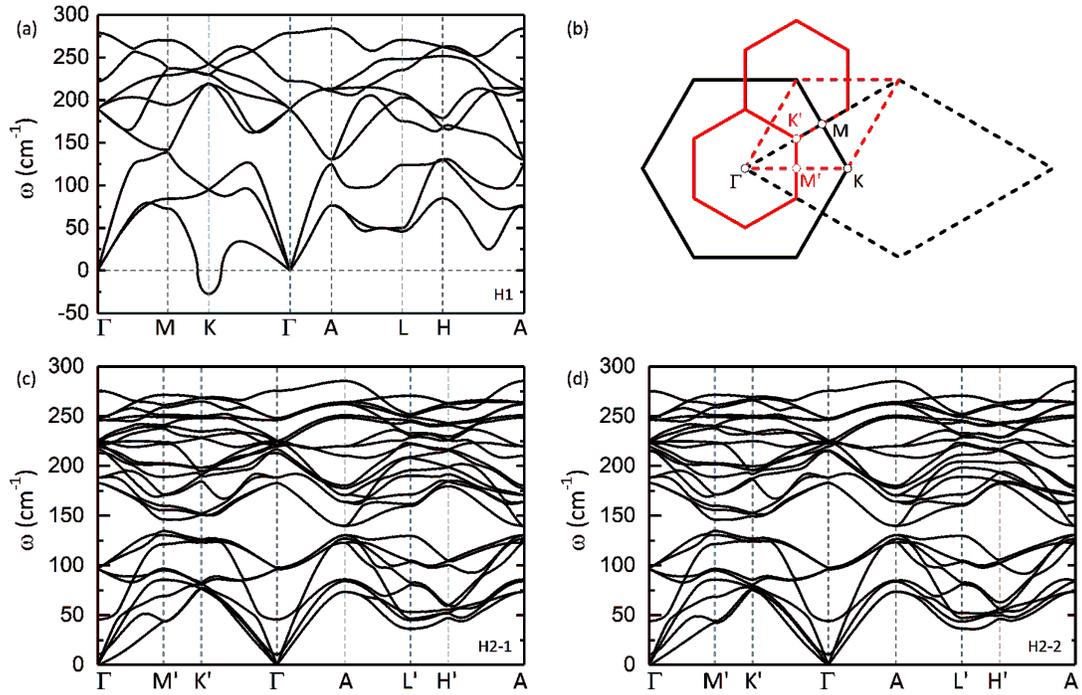

FIG. S2. (Color online) (a) Phonon dispersion of the H1 structure at 30 GPa in the $1 \times 1 \times 1$ unit cell. (b) First Brillouin zones in the $1 \times 1 \times 1$ unit cell (black lines) and the $\sqrt{3} \times \sqrt{3} \times 1$ supercell (red lines). (c)-(d) Phonon dispersions of the H2-1 and H2-2 structures at 30 GPa in the $\sqrt{3} \times \sqrt{3} \times 1$ superlattice.

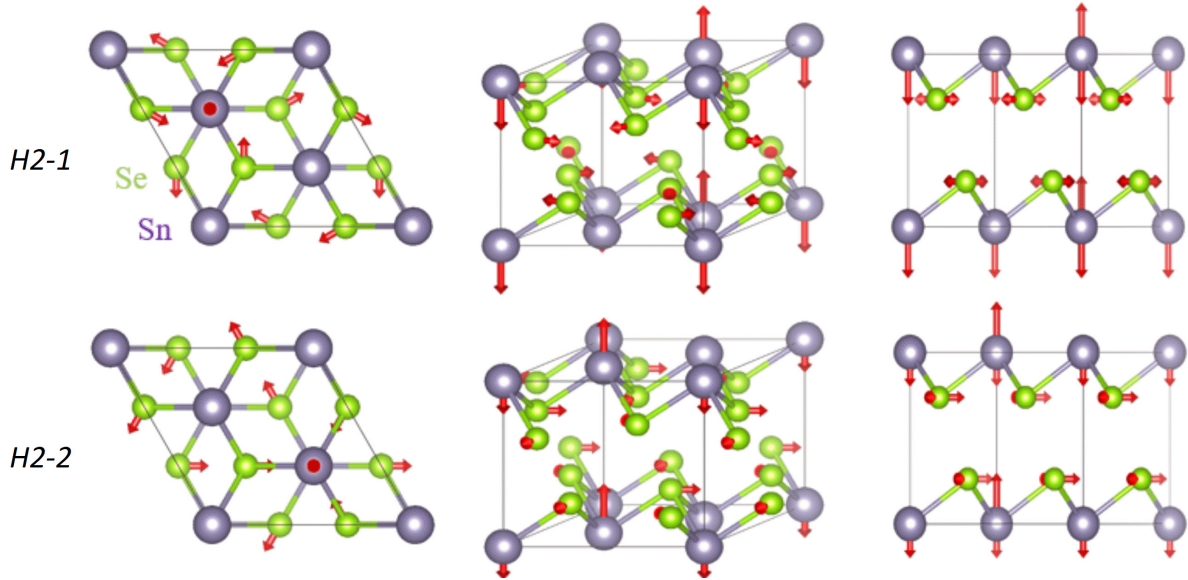

FIG. S3. (Color online) Atomic displacements in the $\sqrt{3} \times \sqrt{3} \times 1$ H1-supercell for the two soft phonon modes leading to the formation of the H2-1 and H2-2 phases. The length of each arrow is proportional to the magnitude of the atomic displacement.

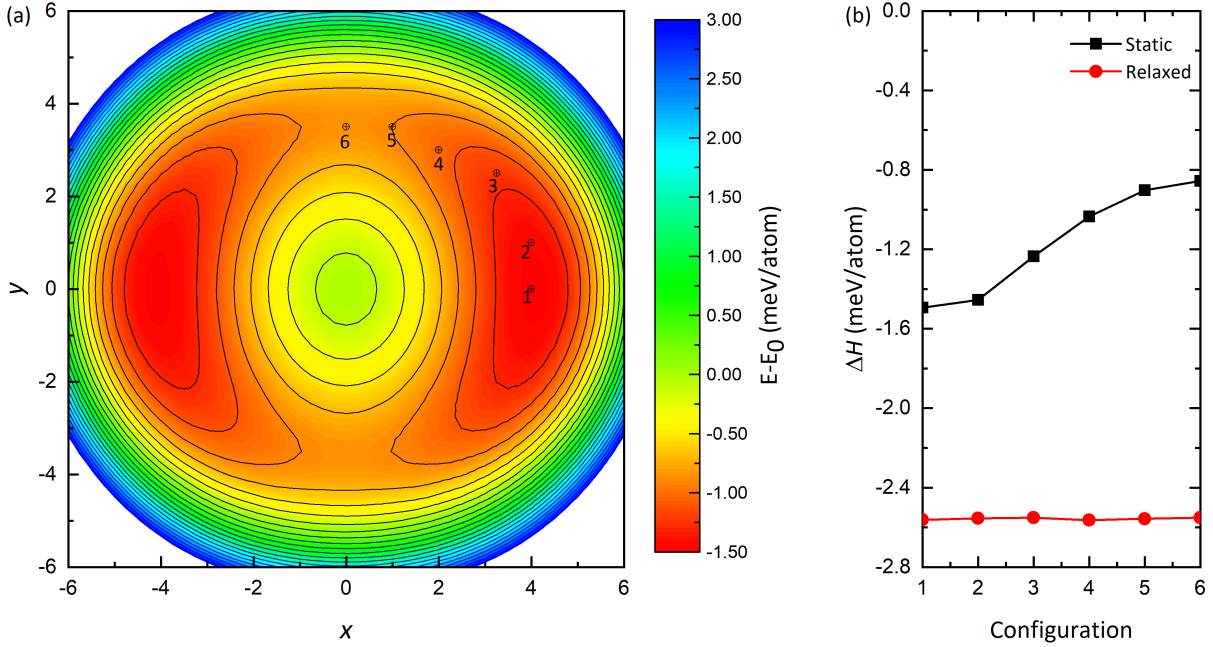

FIG. S4. (Color online) (a) Adiabatic potential energy surface of SnSe$_2$ in the $\sqrt{3} \times \sqrt{3} \times 1$ H1 structure at 30 GPa, calculated along the normal mode coordinates $x, y$ corresponding to the two soft phonon modes at $\Gamma$. The atomic displacements $\Delta\tau_{\kappa,\alpha}$ in direction $\alpha$ are given by $\Delta\tau_{\kappa,\nu}(x,y) = \sqrt{\frac{M_0}{M_\kappa}} \left( e^{\nu_1}_{\kappa,\alpha} x + e^{\nu_1}_{\kappa,\alpha} y \right)$, where $M_\kappa$ is the nuclear mass of atom $\kappa$, $M_0$ is the proton mass, and $e^{\nu}_{\kappa,\alpha}$ are the eigenvectors of the two soft phonon modes. The contour contained many local minima within $\sim 1$ meV/atom energy from each other shown in Fig. S4(a). Full geometrical optimization was performed for the six labeled points on the energy surface. Configuration 1 corresponds to the displacements along only phonon mode $\nu_1$ (i.e. $y = 0$), configuration 6 corresponds to the displacements along only phonon mode $\nu_2$ (i.e. $x = 0$), and configurations 2 to 5 correspond to the displacements where both phonon modes are taken into account. (b) Calculated relative enthalpy with respect to the undistorted phase for the six configurations labeled in (a). The fully relaxed results are shown with red symbols.

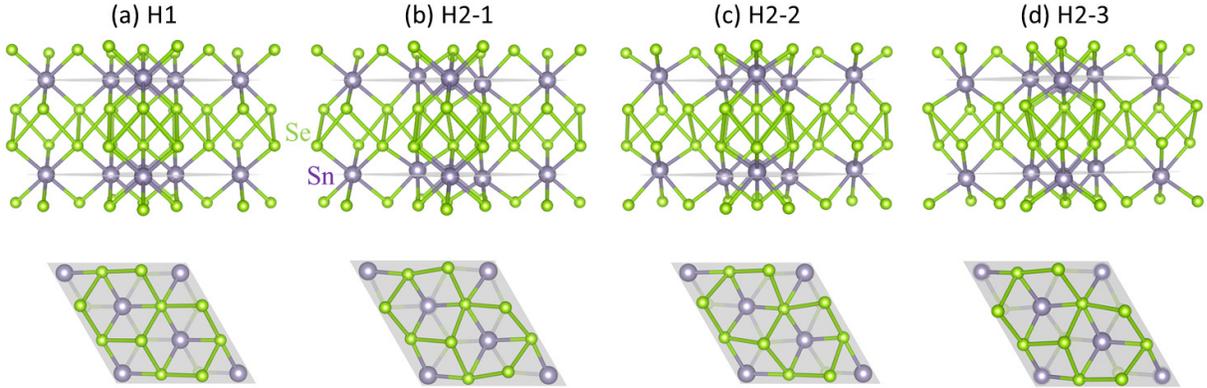

FIG. S5. (Color online) Comparison of the SnSe$_2$ configurations, fully optimized with DFT at 30 GPa. Compared to the parent H1 structure, there is an out-of-plane displacement of the Sn atoms resulting in a slight buckling of the Sn layers and a shift of the Se atoms with respect to the center of the octahedron.

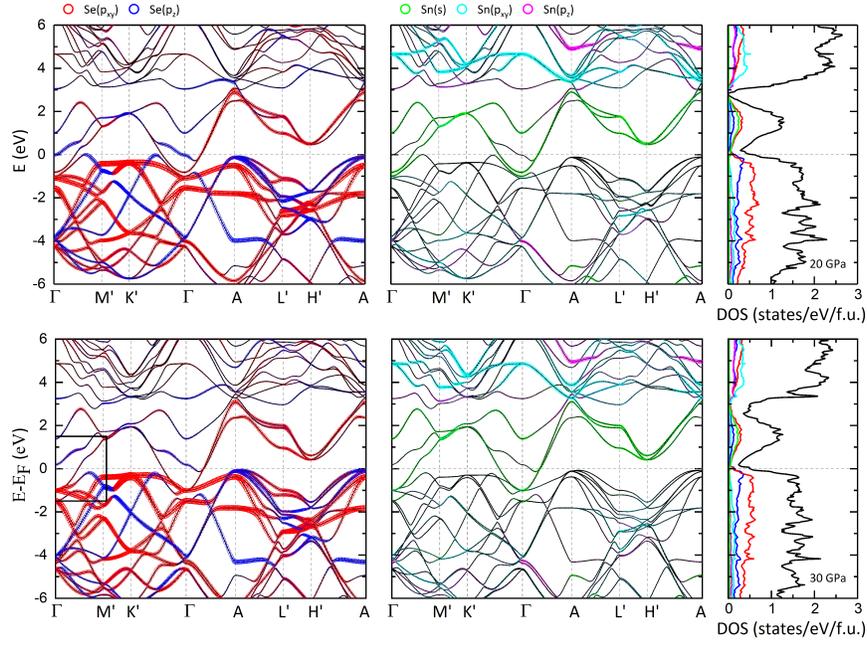

FIG. S6. (Color online) Calculated band structure with orbital characters for the H2-1 structure at 20 and 30 GPa in the $\sqrt{3} \times \sqrt{3} \times 1$ superlattice along a high-symmetry path in the Brillouin zone. The size of the symbols is proportional to the contribution of each character. Avoided crossing near the Fermi level can be observed in the area indicated by the black box (to be compared to the H1 structure in Fig. S11).

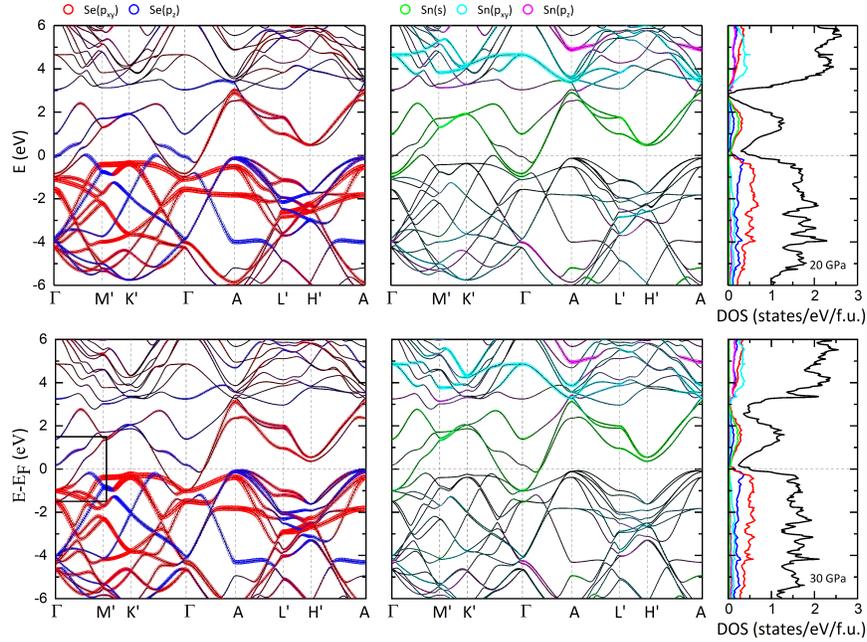

FIG. S7. (Color online) Calculated band structure with orbital characters for the H2-2 structure at 20 and 30 GPa in the $\sqrt{3} \times \sqrt{3} \times 1$ superlattice along a high-symmetry path in the Brillouin zone. The size of the symbols is proportional to the contribution of each character. Avoided crossing near the Fermi level can be observed in the area indicated by the black box (to be compared to the H1 structure in Fig. S11). A comparison with Fig. S6 shows that the differences in the electronic properties of the H2 derivatives are negligible.

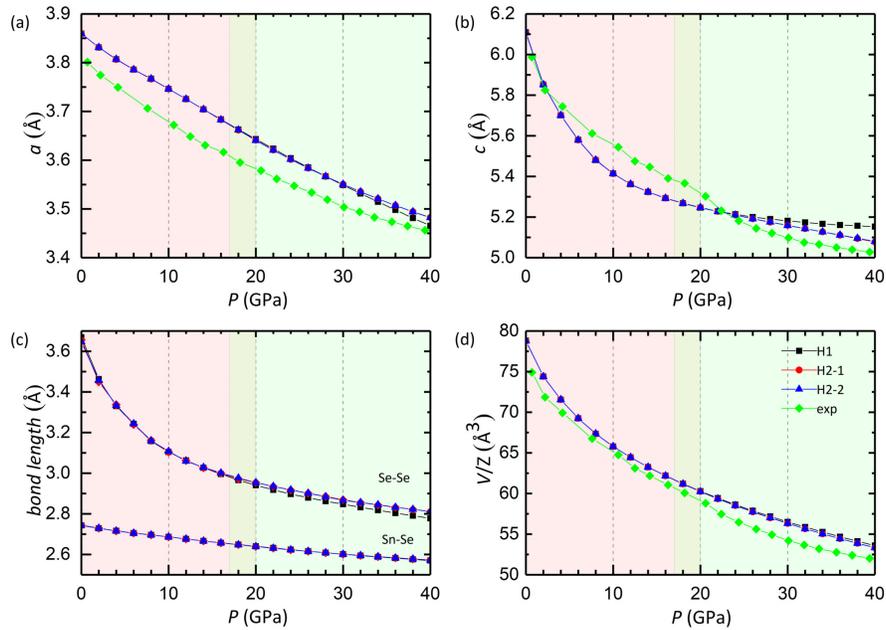

FIG. S8. (Color online) (a)-(d) Pressure dependence of the lattice parameters $a$ and $c$, the average bond lengths, and the volume per one formula unit $V/Z$. The lattice parameter $a$ in the H2 phases was divided by $\sqrt{3}$ to compare with the smaller H1 phase. The in-plane lattice parameter $a$ decreases monotonically and follow the experimental slope as the pressure is increased up to 40 GPa, without displaying any anomaly at the phase transition. On the other hand, the inter-plane lattice constant $c$ drops sharply at low pressures, before beginning a slower decrease around $P \sim 20$ GPa.

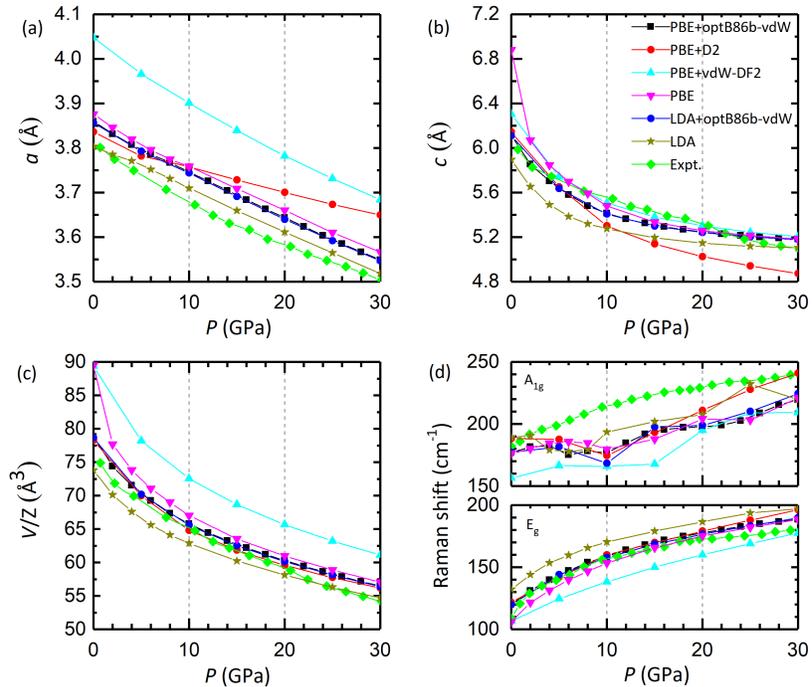

FIG. S9. (Color online) (a)-(c) The effect of the functional choice the lattice parameters $a$ and $c$, the volume per one formula unit, and the frequency of the Raman active modes of H1-SnSe$_2$. Calculations were performed with PBE, PBE with vdW-DF2 [21], PBE with DFT-D2 empirical correction [22], local density approximation (LDA) [23, 24] and LDA with optB86b-vdW [8, 9]. PBE with optB86b-vdW has the best performance, providing lattice parameters in close agreement with the experimental values.

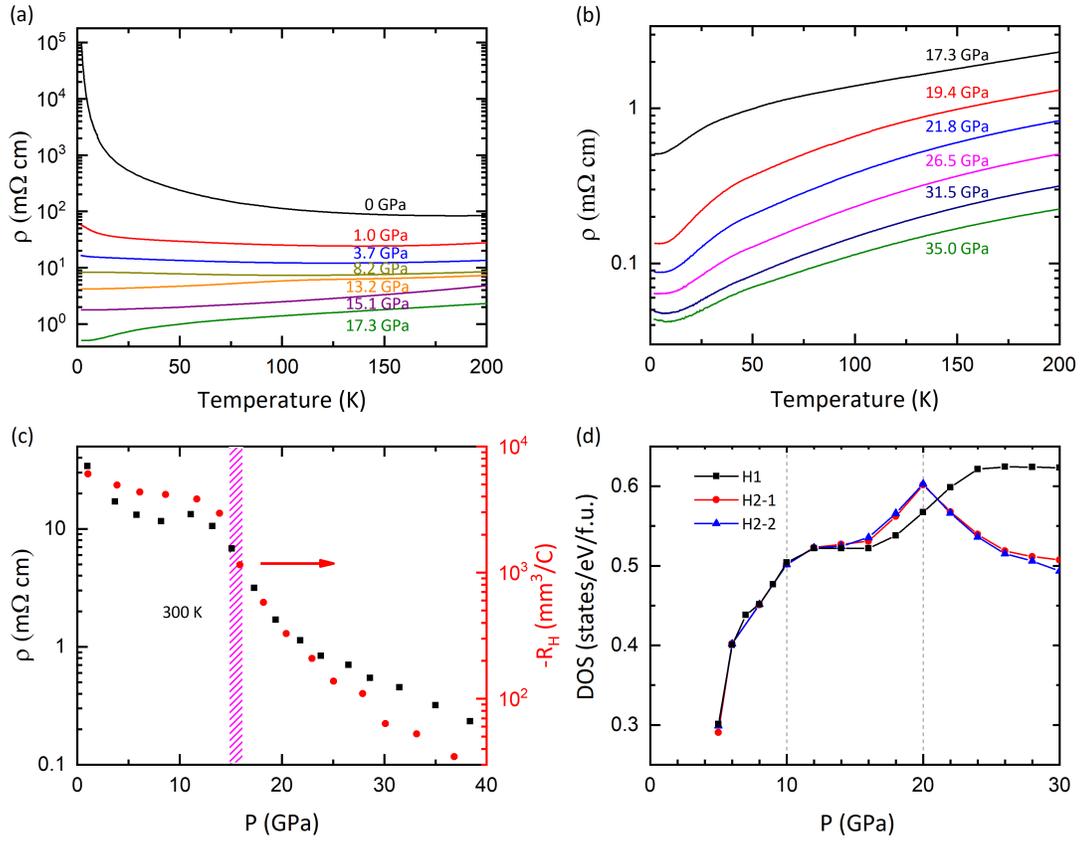

FIG. S10. (Color online) (a)-(c) The measured temperature dependence of the resistivity for various pressures. (a) We observe that the semiconducting behavior is gradually suppressed under pressure. (b) The resistivity becomes metallic and keeps decreasing with increasing pressure in the H2 phase. (c) The pressure dependence of the Hall coefficient (red circles, right axis) and resistivity (black squares, left axis) at 300 K under pressure. (d) The calculated pressure dependence of the density of states at the Fermi level in the H1 and H2 phases, showing a marked decrease for the H2 phases once the transition point has been passed.

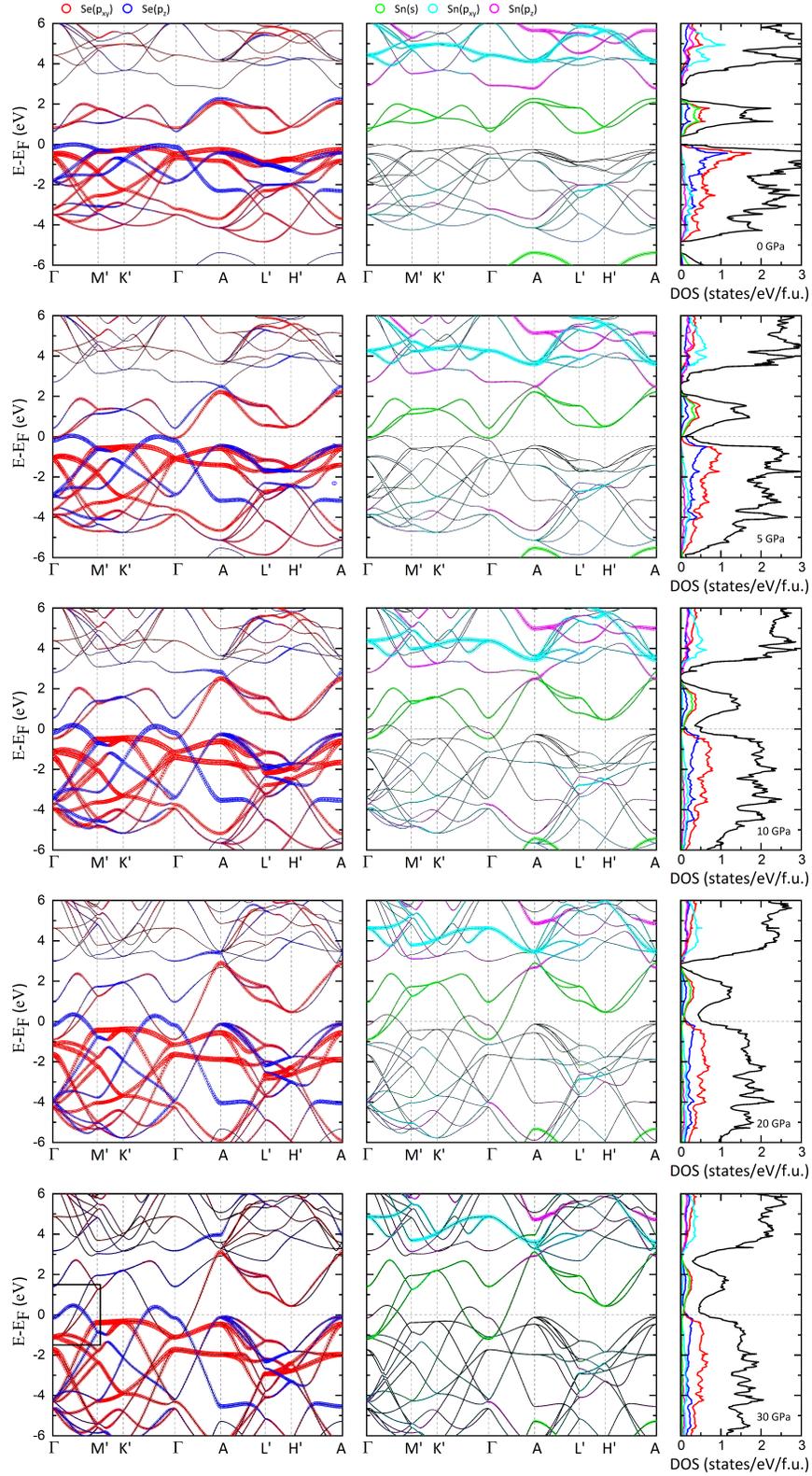

FIG. S11. (Color online) Calculated band structure with orbital characters for the H1 structure at 0, 5, 10, 20, and 30 GPa in the $\sqrt{3} \times \sqrt{3} \times 1$ superlattice along a high-symmetry path in the Brillouin zone. The size of the symbols is proportional to the contribution of each character. As mentioned in the main text, we observe a closing of the band gap for pressure higher than 5 GPa. The area indicated by the black box is to be compared to the Figs. S6 and S7 where avoided crossing near the Fermi level can be observed in the H2 structures.

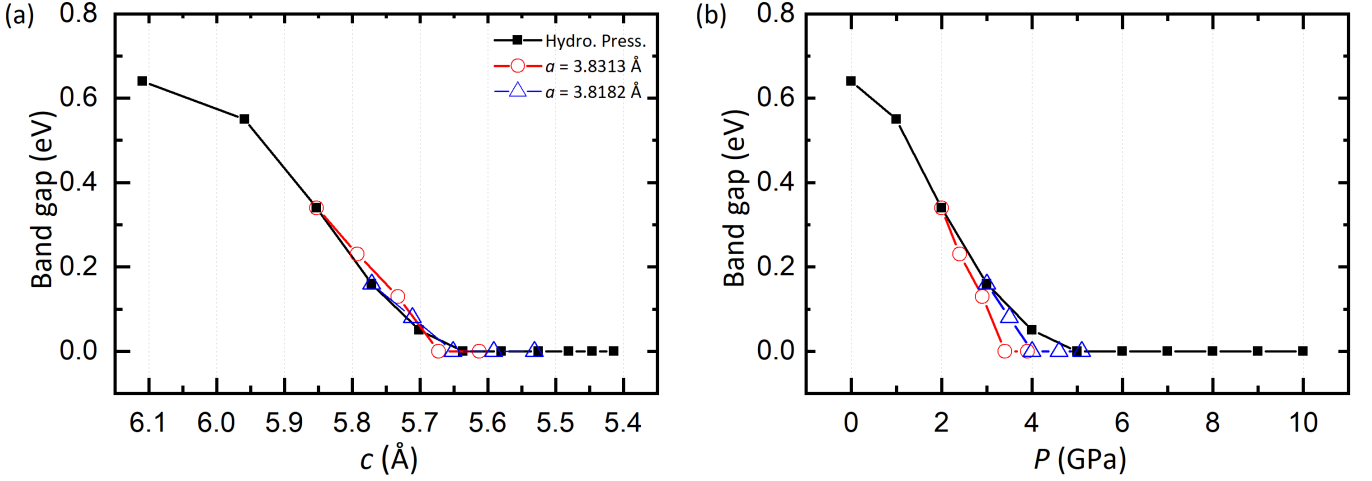

FIG. S12. (Color online) The effect of non-hydrostatic pressure on the electronic band gap. (a) Variation of the band gap as a function of compression of the lattice constant $c$. The lattice constant $c$ was compressed by 1% to 4%, while the lattice constant $a$ was kept fixed at the hydrostatic value from 2 GPa (red circles) and 3 GPa (blue triangles), respectively. Note that the hydrostatic data points at 2 GPa and 3 GPa are also shown in the data curves for compressed $c$ (first point at the left in each curve). (b) Same as in (a) only that the variation in the band gap is shown as a function of pressure.

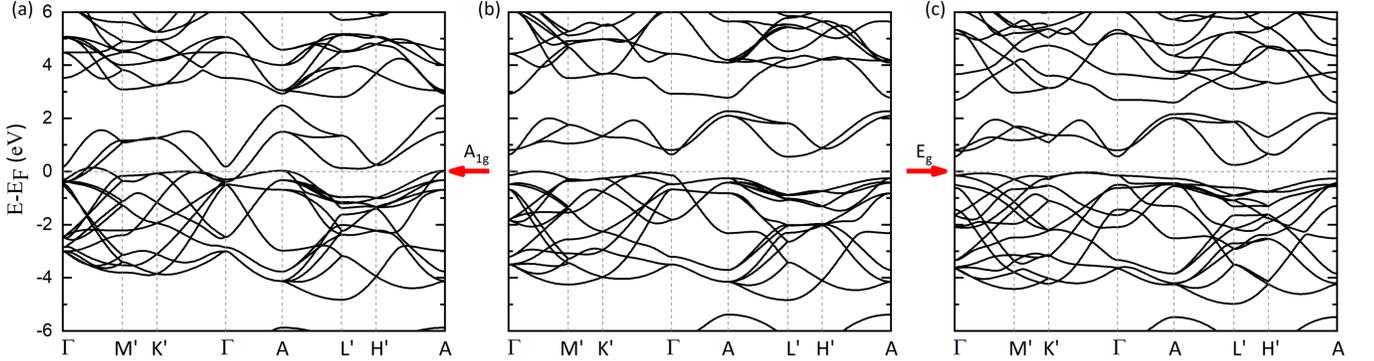

FIG. S13. (Color online) Calculated band structure for the H1 structure at 0 GPa in the $\sqrt{3} \times \sqrt{3} \times 1$ superlattice. In (b) we show the band structure in the ground state, in (a) the band structure calculated after displacing the atoms according to the $A_{1g}$ Raman active phonon mode (out-of-plane displacement of Se atoms with an absolute magnitude of 0.2 Å), and in (c) the band structure after displacing the atoms according to the $E_g$ Raman active phonon mode (in-plane displacement of Se atoms with an absolute magnitude of 0.2 Å). The atomic displacement corresponding to the $A_{1g}$ phonon mode induces large changes in the electronic structure at the Fermi energy indicative of a strong electron-phonon coupling.

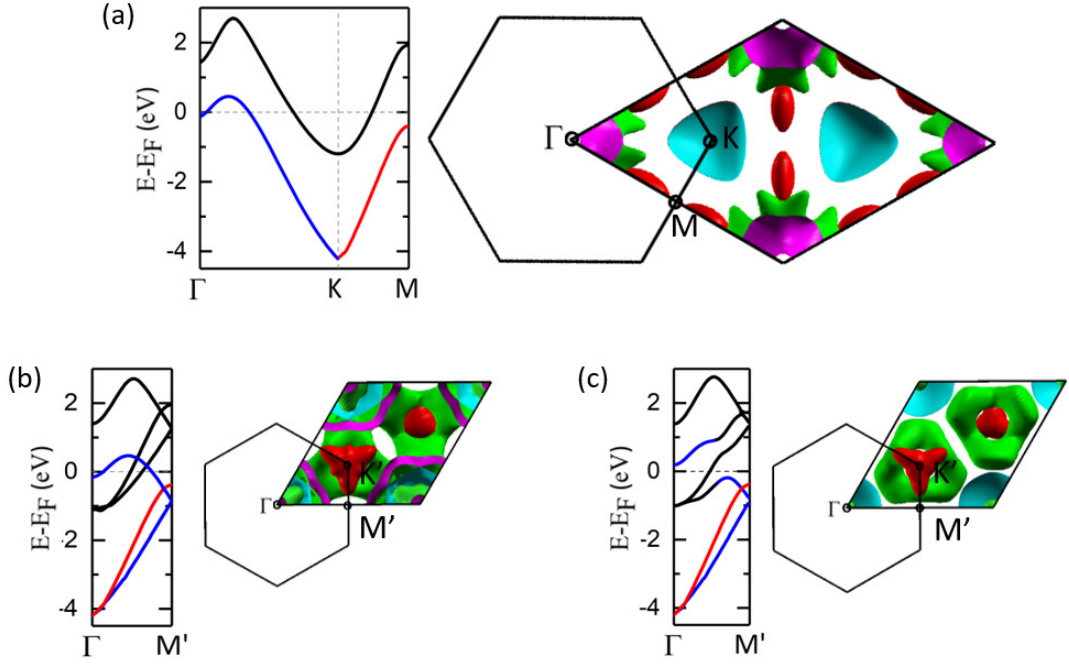

FIG. S14. (Color online) Schematic evolution of the band structure and corresponding Fermi surfaces in the reciprocal unit cell with the phase transition in SnSe$_2$ at 30 GPa: (a) undistorted $1 \times 1 \times 1$ structure, (b) undistorted $\sqrt{3} \times \sqrt{3} \times 1$ superstructure, and (c) distorted $\sqrt{3} \times \sqrt{3} \times 1$ superstructure. The colors of the bands are guides to the eye to allow for easier comparison of the unfolded (a) and the folded (b) and (c) electronic structures.